\documentstyle [11pt]{article} 
\def\n {\noindent}

\def\cle{~\rlap{$<$}{\lower 1.0ex\hbox{$\sim$}}}
\def\cge{~\rlap{$>$}{\lower 1.0ex\hbox{$\sim$}}}

\begin{document}

\begin{center}

\Large

\bf{The Contribution of Late-type/Irregulars to the Faint Galaxy Counts from
HST\footnote{Based on observations with the NASA/ESA {\it Hubble Space
Telescope} obtained at the Space Telescope Science Institute, which is operated
by AURA, Inc., under NASA Contract NAS 5-26555.} Medium Deep Survey Images}

\vspace{0.5cm}

\large Simon P. Driver, Rogier A. Windhorst

Department of Physics and Astronomy, Arizona State University,

Tempe, AZ 85287-1504

\vspace{0.5cm}

and

\vspace{0.5cm}

Richard E. Griffiths

Bloomberg Center for Physics and Astronomy, The Johns Hopkins University,

Baltimore, MD 21218-2695

\vspace{1.0cm}

\vspace{1.5cm} 

Accepted for publication in:

\vspace{0.5cm}

{\it The Astrophysical Journal}, 453 (1995), 48---64.

\vspace{1.5cm}

\end{center}

\normalsize

\pagebreak 

\begin{center} \section*{ABSTRACT} \end{center}

We present a complete morphologically classified sample of 144 faint field
galaxies from the HST Medium Deep Survey with $20.0\leq m_{I}<22.0$ mag. We
compare the global properties of the ellipticals, early and late-type spirals,
and find a non-negligible fraction (13/144) of compact blue [$(V-I)\le 1.0$
mag]
systems with $r^{1/4}$-profiles. We give the differential galaxy number counts
for ellipticals and early-type spirals independently, and find that the data
are
consistent with no-evolution predictions based on conventional flat Schechter
luminosity functions (LF's) and a standard cosmology.

Conversely, late-type/Irregulars show a steeply rising differential number
count
with slope $(\frac{\delta log N}{\delta m}) = 0.64\pm 0.1$. No-evolution models
based on the Loveday {\it et al.} (1992) and Marzke {\it et al.} (1994b) {\it
local} luminosity functions under-predict the late-type/Irregular counts by 1.0
and 0.5 dex, respectively, at $m_{I} = 21.75$ mag. Examination of the
Irregulars alone shows that $\sim 50$\% appear inert and the remainder have
multiple cores.
If the inert galaxies represent a non-evolving late-type population, then a
Loveday-like LF ($\alpha\simeq -1.0$) is ruled out for these types, and a LF
with a steep faint-end ($\alpha\simeq -1.5$) is suggested. If multiple core
structure indicates recent star-formation, then the observed excess of faint
blue field galaxies is likely due to {\it evolutionary} processes acting on a
{\it steep} field LF for late-type/Irregulars. The evolutionary mechanism is
unclear, but 60\% of the multiple-core Irregulars show close companions. To
reconcile a Marzke-like LF with the faint redshift surveys, this evolution must
be preferentially occurring in the brightest late-type galaxies with
$z$\cge 0.5 at $m_{I} = 21.75$ mag.

\vspace{0.25cm}

\n {\it Subject headings:} galaxies: elliptical --- galaxies: spiral ---
galaxies: irregular --- galaxies: luminosity function --- galaxies: evolution


\bigskip

\section{Introduction}

The majority of explanations for the excess of faint blue galaxies (``FBG's'',
see Broadhurst, Ellis \& Shanks 1988, hereafter BES) observed in deep
ground-based CCD images ({\it e.g.} Tyson 1988; Lilly {\it et al.} 1991; Driver
{\it et al.} 1994; Neuschaefer \& Windhorst 1995) involve Irregular/dwarf
\footnote{Here we broadly define dwarf galaxies to have $M_{B} \geq -18.0$ mag
for $H_{o} = 50\ km\ s^{-1}\ Mpc^{-1}$, {\it i.e.} galaxies like the LMC and
fainter.} populations (Kron 1980; Lacey 1991). In some models, these dwarfs
have
rapidly evolved through isolated starbursts (Cowie {\it et al.} 1991; Babul \&
Rees 1992) or via general luminosity evolution (Phillipps \& Driver 1995),
while in other models, strong merging (Broadhurst, Ellis \& Glazebrook 1992;
Rocca-Volmerange \& Guiderdoni 1991) or tidally induced star formation (Lacey
{\it et al.} 1993) are proposed. Other models have suggested that no evolution
is necessary, and that the abundance of FBG's may be due to a combination of
cosmological effects and/or an underestimation of the {\it local} space density
of Irregular/dwarf galaxies (Koo, Gronwall \& Bruzual 1993; Driver \& Phillipps
1995; McGaugh 1994). Whichever process is occurring (and perhaps a combination
of several), one needs high-resolution studies of well a selected galaxy sample
in the magnitude range over which the FBG excess is observed to address their
true nature. Ground-based studies are always limited in resolution due to
atmospheric seeing, because the {\it median} scale-length of faint field
galaxies is $\sim0.3''$ (Griffiths {\it et al.} 1994b, GR94b; Casertano {\it et
al.} 1995, CRGINOW). Giraud (1992) found evidence for both merging, isolated
starbursts, and possibly post-starburst remnants. The studies of Burkey {\it et
al.} (1994) and Colless {\it et al.} (1994) also find a significantly increased
fraction of close companions for $19\leq m_{I}\leq 22$ mag. The Colless {\it et
al.} sample was limited by ground-based seeing to a resolution of
0.5$''$---1.0$''$ FWHM. The superb resolution provided by the refurbished HST
now allows morphological details to be seen to much higher resolution and
fainter limits (Griffiths {\it et al.} 1994a, GR94a; Forbes {\it et al.} 1994;
Glazebrook {\it et al.} 1995a, GL95a) and in particular allows us to study the
morphology, light-profiles and contours of individual galaxies to a resolution
of $\sim 0.1''$ FWHM in the flux interval over which the excess FBG's are
observed ({\it i.e.} $m_{B}\geq 22$ mag; BES, Driver {\it et al.} 1994). Here
we
present a complete sample of HST field galaxies with $I$-band magnitudes in the
range $ 20.0\leq m_{I}<22.0$ mag (or $22.0<m_{b}<23.5$), from which we extract
a complete sub-sample of galaxies that are irregular in appearance. In \S 2 we
summarize the new HST WFPC2 data and its method of data reduction. In \S 3 we
present the detection and photometry algorithms used to construct the current
catalog. In \S 4 we discuss the morphological classification, and in \S 5 we
compare the global properties of each galaxy class. In \S 6 we present a
discussion of the detailed morphology of the Irregular HST galaxies.

\section{The HST WFPC2 data}

The HST Medium Deep Survey (MDS) collects data in parallel with other HST
instruments (FOC, FOS, FGS) using the WFPC2 to randomly image a field
4$'$---14$'$ away from the primary target (depending on the primary HST
instrument being used). A more detailed description of the MDS project is given
by GR94b, and a summary of the Cycle 4 HST data reduction methods is given by
Ratnatunga {\it et al.} (1995). The MDS pipeline data reduction was carried out
using the MDS WFPC2 super-skyflats (Ratnatunga {\it et al.} 1994). This results
in a uniform calibration accuracy of $\pm 0.1$ mag (Holtzman {\it et al.}
1995).
The actual zero magnitude points for 1.0 ADU/sec are: $K_{I_{814W}} = 21.67 -
0.009 (F606W-F814W)$ mag and $K_{V_{606W}} = 22.84 - 0.076 (F555W-F814W)$ mag.
(Note that $(F606W-F814W)\simeq$1.4 and $(F555W-F814W)\simeq$1.8 mag for faint
field galaxies, GR94a, GR94b). The fields were selected from the MDS database
to have comparable exposure times in both the wide $V$ (F606W) and $I$ (F814W)
filters. Table 1a shows the positions and exposure times for the data used in
this study, along with Galactic foreground extinction in $V_{606}$ and
$I_{814}$, taken from the reddening values in Burstein \& Heiles (1982). We
adopted an extinction $\propto 1/\lambda$ (Osterbrock 1989) and a ratio of
$A_{V}/E(B-V) = 3.2$. The removal of cosmic rays was achieved using the
CRREJECT
facility in IRAF's IMCOMBINE, which compares individual orbits to reject
cosmic-rays and creates a final stack via a sigma-weighted average. In general,
we did not have a sufficient number of undithered orbits per field to apply the
optimized CR rejection routines of Windhorst, Franklin \& Neuschaefer (1994).
Additional low-level cosmic rays and bad pixels were therefore cleaned by
replacing pixel values 3$\sigma$ above or below the local median sky-background
with the mean from adjacent pixels. The WFC pixel scale is 0.0996$''$ per
pixel, and so each of the three WFC CCDs corresponds to a $1.3'\times 1.3'$
region.\footnote{There is a potential bias introduced by the position of the
primary HST target and the desire for long integrations. However, most primary
targets are themselves randomly detected ({\it i.e.} QSO's or stars at
relatively high galactic latitude), so any such bias is expected to be small.}
This gives a total sky coverage of 0.00845 sq deg. Data from the PC was not
used in this survey due to its poorer surface brightness (SB) sensitivity and
its much lower sky coverage.

\section{A complete catalog of Irregular HST galaxies to $m_{I} = 22$ mag}


In order to define a complete sample of Irregular HST galaxies, and study their
morphology in an unbiased way, it was necessary to first construct a complete
sample for {\it all} galaxy types from the MDS images. We therefore required
that the magnitude limit was not so faint that the Irregulars become
unresolvable by HST. Typical field Irregulars are expected in the range
20.0\cle $m_{I}$\cle 22.0 mag and 0.5$''$\cle $r_{half-light}$\cle 1$''$ for
$z\leq 0.3$,
assuming --14.5\cle $M_{I}$\cle --19.5 mag, $(B-I)\simeq$1.5, and $r_{hl}\simeq
1-3$kpc (for $H_{o} = 50\ km\ s^{-1}\ Mpc^{-1}$, $q_{o}=0.5$, and $\Lambda=0$).
This observed range is equivalent to a $B$-band magnitude range of
$22.0<m_{B}<23.5$ (Driver {\it et al.} 1994), at which level the observed
density of FBG's is a factor of $\sim 2-3$ over the standard cosmological
predictions (Tyson 1988; see our Figures in \S 5). The expected surface density
of galaxies of {\it all} types in this $I$-band range is $\sim 20$ objects per
WFPC2 field (Tyson 1988; Driver {\it et al.} 1994; Table 1b here).

\subsection{Image detection and photometry}

Prior to the image detection phase, a median filter of $127\times 127$ pixels
was used to create a smoothed sky-image for each WFPC2 frame, representing any
large-scale variations in the sky (gradients of $\sim 2$\% typically remained
after applying the MDS supersky flats). This sky-image was subtracted from the
original frame to give an extremely flat sky-background ($<<2$\%). We divided
each CCD frame into nine sections and measured the sky background within each
section, using a Gaussian fit to the peak of the ADU-histogram, which reduced
the WFPC2 gradients of 2--3\% to \cle 0.5\% of sky.

The initial object detection was done with the isophotal detection routine {\sc
`IMAGES'} in the {\sc RGASP} package (Cawson 1983), searching for objects with
4 connected pixels (0.04$''^{2}$) with a signal 2 $\sigma$ above the mean
sky-background (where $\sigma$ is the measured noise in the {\it local}
sky-background). Detections with centroids within 26 pixels (2.6$''$) of the
CCD
frame edge were rejected to prevent any edge bias. This reduces the usable area
per CCD image to 660$^{2}$ pixels or $1.1'\times 1.1'$. For the photometry
stage
we used a variable size circular aperture centered on the initial detection
position and a global sky-background for each CCD image (because any
significant
large-scale structure in the sky background was already removed to $\leq 0.5$\%
or $\sim$28.5 mag/arcsec$^{2}$). The size of the aperture, $r_{ap}$, was based
on the following relationship:

\begin{equation} r_{ap}^{n}=r_{iso}^{n}+r_{min}^{n}, \end{equation}

\n as described in Driver (1994) and in Jones {\it et al.} (1995). This
equation
essentially satisfies the conditions that the aperture radius is comparable to
the isophotal radius $r_{iso}$ for objects with large isophotal areas ({\it
i.e.}, bright), and comparable to a {\it fixed} minimum $r_{min}$ for objects
with small isophotal radii ({\it i.e.}, faint). The following optimal values
for
pure exponential disks were determined via simulations (Driver 1994): n=1.5 and
$r_{min}\simeq 1''\simeq$3 scale-lengths (CRGINOW). These simulations showed
that this technique includes at least 95\% of the light for exponential disks
as
well as bulge-dominated galaxies. All detections with an aperture flux
$m_{I}\leq 22.0$ mag were re-evaluated individually to exclude any
contamination
due to nearby objects within the initial aperture. The final catalog was then
defined to be all galaxies within the {\it total} flux range of $20.0\leq
m_{I}<22.0$ mag.

\subsection{Completeness of the HST WFPC2 sample}

Figure 1 shows the logarithm of the signal-to-noise ratio versus $I$-band
magnitude for the isophotal detection stage (Fig. 1a) and the photometry stage
(Fig. 1b) as a test of our sample completeness. The majority of detected
galaxies are well above the signal-to-noise limit set by the detection
criterion
({\it i.e.} 2$\sigma$ above sky over 4 contiguous pixels). Fig. 1b also follows
the predicted slope of $-0.6$ in a Euclidean Universe (shown as dashed line)
very closely for optimal aperture radii ({\it i.e.} the radius which just
encompasses the object). Galaxies with central SB $\mu_{I}\ge 23.26$ mags
arcsec$^{-2}$ ({\it i.e.}, whose extent is greater than 5$''$ in radius, see
Table 1b) would escape detection. Since no objects are seen close to the quoted
signal-to-noise limit ($\simeq$4.0 in Fig. 1a), the sample is essentially 100\%
complete down to a limiting {\it total} magnitude of $m_{I} = 22.0$ mag. A
discussion of the completeness of WFC images much closer to the HST detection
limit is given by Neuschaefer {\it et al.} (1995). Note that the total {\it
volume} surveyed for low-SB galaxies is still very small due to the limited
WFC2
field-of-view. The catalog may therefore be statistically mis-represented for
these types (Disney 1976). Table 1b shows the number of detections per WFPC2
CCD along with the magnitude zero points and SB detection limits. The number
of detections agrees well with that expected (\S 5.1).

\subsection{Morphological Classification of the WFPC2 galaxies}

The morphological classification of the HST galaxies was achieved from the
consensus of three independent eyeball classifiers (SPD, RAW and Roger Rouse -
RR). To assist in the classification, greyscale plots were made from $1\sigma$
below sky to $5\sigma$ above sky, as well as major-axis profiles in both SB vs.
r and SB vs. r$^{\frac{1}{4}}$. These were used to classify all objects in {\it
both} $V$ and $I$. Simulated light-profiles of perfect de Vaucouleur-laws and
exponential disks were also plotted on the same scales as a reference. The
following rules were followed for the classification of our HST galaxies, based
on the Hubble Atlas of Galaxies (Sandage 1961):

\n {\bf (1)} Visually compact objects showing a predominately linear profile in
SB vs. $r^{\frac{1}{4}}$ were classified as {\it E/S0}. This includes compact
systems which show no evidence for a disk, {\it irrespective} of their colors.

\n {\bf (2)} Objects which are linear in $r^\frac{1}{4}$ in the inner region
and linear in r in the outer region were classified as {\it Sa, Sb or Sc},
depending
on the ratio between the linear sections and the steepness of the respective
profiles (see e.g., Windhorst 1994a, 1994b). Note that the Sabc's are lumped
together into a single galaxy class for the purpose of classifications in this
paper. This class thus includes those systems that have a well defined bulge
and disk.

\n {\bf (3)} Objects with a flat light-profile, a profile which rises from the
center, an erratic profile, or a profile which is a poor linear fit to either
SB vs. r or SB vs. r$^{\frac{1}{4}}$ were classified as {\it Sd/Irregular or
Peculiar}. Peculiar galaxies are those which exhibited a strong central core
(initially linear in SB vs. r$^{\frac{1}{4}}$), but with a highly disturbed
outer disk. This is an attempt to distinguish between genuine Irregulars and
early-type galaxies undergoing some form of interaction, etc.

Note that major-axis light-profiles were used rather than azimuthally-averaged
profiles to prevent any irregular structure being smoothed out. In all cases,
the grey-scales plots were used as a secondary check to determine whether a
nearby companion might be disturbing the light-profiles. Both the $V$ and $I$
light-profiles {\it and} grey-scale plots were used for optimal classification.
Figure 2 shows a selection of major-axis profiles to illustrate the range of
Hubble types and light-profiles seen in our HST sample. A representative galaxy
for each Hubble class was chosen with $m_{I}\sim 21.5$ mag to allow a direct
comparison of the appearance and light-profiles of the fainter objects.

Comparing the results between the independent classifiers, we find complete
agreement for 91/144 objects and agreement to one Hubble class for 45/144
objects. The remaining 8 objects were typically unusual cases where one
classifier identified the major component of a clump and the other classified
the entire system as ``Peculiar'' or Irregular. In these cases the
classification of the objects was openly discussed, and a consensus
classification was assigned. No attempt was made to separate stars from compact
ellipticals, as the two populations sometimes appear indistinguishable by
visual examination of their HST light profiles (see Ratnatunga {\it et al.}
1995 for an
automated method of star-galaxy separation). Currently, a full spectroscopic
follow-up of the entire sample is in progress (Driver {\it et al.} 1995b) to
determine redshifts and luminosities. This will allow an accurate separation
between compact ellipticals and stars, and enable us to confirm the eyeball
classifications and determine how many high-SB galaxies are masquerading as
stars at faint magnitudes.

\subsection{Reliability of the WFPC2 Classifications}

Figure 3 shows a simple test of our classification by plotting the
concentration index (C.I. = core {\bf SB} minus total magnitude) versus total
magnitude. Here,
the core SB is defined as the flux within 0.2$''$ radius in units of magnitudes
per square arcsecond. The C.I. is a measure of the concentration of galaxy
light, where a lower index reflects a greater amount of light contained in the
core, which is expected to correlate well with morphological type (Forbes {\it
et al.} 1994). This should not be confused with the more conventional
definition of the C.I. (e.g. Kent 1985), which requires better resolution and
signal-to-noise than we have. In reality, the core SB is both resolution and
distance dependent. However at moderate redshifts the $\Theta-z$ relation for
HST bulges and disks is flat enough (Mutz {\it et al.} 1994) that the C.I. is
to first order distance independent.

Figure 3 shows that the E/S0 sample ({\it i.e.} compact systems including a
small number of stellar-like objects) separates very well from the bright
spirals. However, the transition from early to late-type spirals is less
well-defined by this technique, although there is still a distinct correlation
between galaxy type and the C.I..


As a final check we compare our morphological classifications to those of GL95a
and GR94a. Both of these surveys are also based on HST MDS data. The survey of
GL95a covers several fields in common with this study, yet uses entirely
independent methods for data reduction and image detection. Comparing the
classifications for the 92 galaxies in common, we find that 25 agree exactly,
44 within one Hubble class, and 8 within two Hubble classes. The remaining 14
objects are unusual in some way, and represent cases where a decision had to be
made as to whether to classify the major component of a clump or the entire
system. In their magnitude limited sample, GR94a find a mixture of 19\% E/S0,
44\% Sa-Sm, 13\% Irregulars/Mergers, and 25\% Peculiar/Unclassified, whereas we
find: 32\% E/S0, 53\% Sa-Sd, and 15\% Irr/Peculiar. The agreement clearly
depends on the distribution of the Peculiar/Unclassified class of GR94a amongst
our classes, but they note that a significant number ({\it i.e.} about half) of
their ``unclassified objects'' were classified as S0's by one of their two
classifiers. If so, this would suggest very good agreement between the two
samples. Overall, we consider the agreement between this study and those of
GL95a and GR94a to be good, providing a consistent picture of the field galaxy
mix at faint magnitudes.

Nevertheless, we recognize the need to develop an automated classifier for
faint
field galaxies. Given the good agreement with the other independent eyeball
classifiers of GL95a, we consider our sample to be a suitable representative
control set and thus a good training set for Neural Network classifiers. Such
automated classifications methods are essential, as the MDS database of WFPC2
images has by now accumulated over 200 fields containing several thousand
galaxies. Finally, we note that a full spectroscopic survey is underway, which
will help confirm the reliability of these classifications, although we point
out that faint E/S0's and Sa's, as well as faint Sb's and Sc's, are hard to
distinguish spectroscopically at moderate redshifts (Keel \& Windhorst 1993;
Windhorst {\it et al.} 1994a, b).

\subsection{The WFPC2 Galaxy Catalog down to I$\le$22 mag}

Table 2 shows the full WFPC2 catalog which contains the basic parameters for
the 144 MDS field galaxies from which we draw our Irregular galaxy sample. The
random error in the listed magnitudes is $\pm 0.06$ mag. The largest error
comes from the limited accuracy in the sky-subtraction. Given that large-scale
residuals in the sky are of order 0.5\%, an average aperture radius of 16
pixels
(see Column 15 of Table 2) yields random errors of $\sigma_{I} = 0.06$ and
$\sigma_{V} = 0.10$ mag for an object with $I = 22.0$ and $V = 23.3$ mag (and a
mean sky value of $\mu_{I}=21.7$ and $\mu_{V}=22.3$ mag arcsec$^{-2}$). For the
$(V-I)$ color, this implies a random error of $\pm 0.12$ mag.

More information about the objects in Table 2, such as their scale-lengths, is
given by and Ratnatunga {\it et al.} (1995). A discussion of the scale-lengths
of MDS galaxies is also given by Mutz {\it et al.} (1994) as function of
redshift, and by CRGINOW and Im {\it et al.} (1995a) as function of apparent
magnitude.

The classifications in Table 2 are listed according to Hubble class as defined
above. For the remainder of this paper, the Hubble classifications that we
assigned to the individual WFPC2 galaxies were binned into three classes: E/S0
(compact), Sabc (early-disk), and Sd/Irr (late-disk) plus Peculiar galaxies.
The Sd's were grouped together with Irr's so as to obtain roughly equal numbers
in each of the three categories.

\section{Global properties of the Late-type HST galaxies}

Without complete redshift information, we cannot distinguish between genuine
Irregular galaxies of low intrinsic luminosity at lower redshifts and normal
galaxies which have developed an irregular appearance, perhaps due to merging
or a violent asymmetric burst of star formation at moderate to large redshifts.
The
rather low mean redshifts observed in the field galaxy redshift surveys (BES;
Colless {\it et al.} 1990, 1991, 1993; Cowie {\it et al.} 1991; Lilly {\it et
al.} 1991) imply that only limited luminosity evolution can have occurred in
the
intrinsically brighter galaxy populations. Hence, those galaxies with highly
irregular appearance are more likely to be intrinsically low-luminosity systems
with inherent irregular morphology, or intrinsically low-luminosity systems
undergoing some evolutionary process, or a combination thereof. For the
purposes of this paper, however, we exclude the "Peculiar" galaxies and
consider only the ellipticals, early- and late-type spirals plus what we
believe to be the genuine
Irregulars based primarily on their low apparent central SB and flat
light-profiles. The mean apparent central SB for each of these types is:
$\mu_{I}^{E} = 18.5\pm 0.2$, $\mu_{I}^{Sa} = 19.9\pm 0.2$, $\mu_{I}^{Sb} =
20.5\pm 0.2$, $\mu_{I}^{Sc} = 20.8\pm 0.3$, $\mu_{I}^{Sd} = 21.2\pm 0.3$ and
$\mu_{I}^{Irr} = 21.8\pm 0.2$ mag/arcsec$^{2}$. Note that our central SB was
measured as the core SB within the central (0.2$''$) ellipse from the
RGASP PROF package (Cawson 1983). Quoted rms errors are based on the number of
objects binned into each group.

Figure 4 shows the integrated $(V-I)$ color versus magnitude for the full
sample
with the equivalent histogram overlaid. There is little or no distinct trend in
color with apparent magnitude {\it nor} with morphological type. This is mainly
due to the rather small wavelength baseline between the F606W and F814W filters
(which however overlap little in wavelength). {\it No} Irregulars are seen with
colors redder than $(V-I)\simeq 1.5$ mag, and {\it no} spirals exhibit colors
redder than $(V-I)\simeq 2.0$ mag. The shaded area shows the distribution of
Irregulars only, and is not found to be significantly bluer than that of the
overall sample [$(V-I) = 1.2\pm 0.1$ mag]. For each individual type the
observed mean colors are: $(V-I)_{E} = 1.4\pm 0.2$, $(V-I)_{Sa} = 1.2\pm 0.2$,
$(V-I)_{Sb} = 1.1\pm 0.2$, $(V-I)_{Sc} = 1.0\pm 0.3$, $(V-I)_{Sd} = 1.1\pm 0.3$
and $(V-I)_{Irr} = 1.0\pm 0.2$ mag.

Note that the mean color for the entire sample agrees well with that determined
by GR94b when converted to a common filter system (Bahcall, {\it et al.} 1994).
Figure 5 shows the predicted color versus redshift for the various galaxy
types.
The local colors for each Hubble type were taken from the models in Windhorst
{\it et al.} (1994b): $(V-I)_{E/S0}\simeq 1.4\pm 0.2$, $(V-I)_{Sabc}\simeq
1.1\pm 0.2$, and $(V-I)_{Sd/Irr}\simeq 1.0\pm 0.2$ mag. K-corrections for
galaxy
types E/S0, Sa, Sb, Sc, and Sd/Irr were derived from the present day spectra of
Guiderdoni \& Rocca-Volmerange (1987, 1988) for the WFPC2 $V_{606}$ and
$I_{814}$ filters (Myungshin Im 1995, private communication). Based on their
likely zero-redshift SED's, the Hubble classes are expected to have a {\it
range} in observed $(V-I)$ color of $0.9<(V-I)_{E/S0}<2.7$, $0.7 <
(V-I)_{Sabc}<2.2$, and $0.3<(V-I)_{Sd/Irr}<1.8$ mag, depending on their exact
redshift distribution. Our observed $(V-I)$ color range is generally consistent
with these expectations: the only major exception is the surprising abundance
of
very blue E/S0's with $(V-I)$\cle 1 mag. This may imply an error in our method
of classifications ({\it i.e.} a preponderance of Blue Compact Dwarf systems
mis-classified as E/S0's), or that the E/S0 population has evolved (Charlot \&
Bruzual 1991), and/or a strong color-luminosity relation for E/S0 galaxies.
Blue
E/S0 systems are known to exist in other samples studied with HST (e.g. compact
narrow emission-line galaxies, c.f. Koo {\it et al.} 1995, see also Im {\it et
al.} 1995b), but are not normally noted in significant numbers in ground-based
field surveys. However, they could have been missed as E/S0's in typical
ground-based seeing. Our sample of E/S0's has a preponderance of $r^{1/4}$
light-profiles, even though a non-negligible fraction has rather small
scale-lengths ($r_e$\cle 0.3'', CRGINOW). A spectroscopic follow-up is needed
to reveal the nature of these blue E/S0 classifications.

Figure 6 shows the change in morphological mix from the bright to the faint end
of our magnitude range, and shows a distinct increase in the {\it ratio} of
late-type galaxies to bright spirals plus ellipticals. The percentage mix
changes from 36\% E/S0, 50\% Sabc, 14\% Sd/Irr at $m_{I} = 20.25$ mag to 28\%
E/S0, 35\% Sabc and 31\% Sd/Irr at $m_{I} = 21.75$ mag (+6\% Peculiar). There
is thus a rapid increase in the number of galaxies with late-type morphology
towards progressively fainter magnitudes. Our high resolution HST images thus
confirm the initial claims that late-type galaxies are responsible for the
excess of faint blue objects observed in deep CCD surveys (see \S 1). The
bright
end of our sample compares well to that observed by Shanks {\it et al.} (1984),
who found 43\% E/S0, 45\% Sabc and 12\% Sd/Irr. Note that galaxies classified
as
Peculiar ({\it i.e.} obvious mergers or images with a bright core coupled with
irregular structure) also become more prevalent at fainter magnitudes. This may
be a reflection of the increasing volume surveyed at fainter magnitudes making
the catalog more complete for rarer types. Alternatively, it could represent an
epoch at which brighter field galaxies merged more frequently (Burkey {\it et
al.} 1994). Clearly, with only 7 galaxies it is impossible to draw substantial
conclusions for the Peculiar types other than that they {\it may} represent
evolution in a {\it small fraction} (7 of $\sim$ 110) of the brighter field
population, which {\it appears} to become more frequent at fainter magnitudes.

\section{The Morphological Galaxy Counts Observed with HST}

\subsection{The observed counts as a function of galaxy type}

Figure 7 shows the more conventional differential galaxy number counts versus
$I$-band magnitude plots for: (a) the total galaxy sample; (b) the
elliptical/compact galaxies (E/S0's); (c) spiral galaxies (Sabc's); and (d)
late-type galaxies (Sd/Irr). The counts of the total sample are linear,
suggesting that the sample is indeed representative and complete down to at
least I=22 mag, as inferred from Figure 1. However, the slope is slightly
steeper ($0.43\pm 0.05$) than that observed by other groups in the $I$-band
(Tyson 1988; Driver {\it et al.} 1994; Burkey {\it et al.} 1994, Neuschaefer \&
Windhorst 1995). This may suggest that the ground-based samples suffer from
partial incompleteness or star-galaxy confusion (Neuschaefer \& Windhorst 1995,
CRGINOW), or more likely that the discrepancy is a reflection of the limited
magnitude range covered by our high S/N sample (the ground-based slopes are
typically measured out to fainter magnitudes, where the counts are expected to
flatten). The individual counts of ellipticals (E/S0's --- Fig. 7b) and
early-type spirals (Sabc's --- Fig. 7c) have flatter and comparable slopes of
$0.31\pm 0.05$ and $0.34\pm 0.05$, respectively. The counts of the late-type
galaxies, however, exhibit a much steeper and rather unexpected slope of $0.64
\pm 0.1$ (Fig. 7d), consistent with the Euclidean value. GL95a independently
find a similar trend for the fractions of morphological types over a comparable
magnitude range. We conclude that the steep number counts of the
late-type/Irregular population likely gives rise to the faint blue galaxies
observed at faint magnitudes. If this trend continues, this implies that the
faint galaxy number counts (and the Extragalactic Background Light) are largely
if not entirely dominated by late-type galaxies. Are these late-type/Irregulars
expected, are they evolving, and if so through what mechanism?


\subsection{Modelling the Morphological Galaxy Counts}

To compare these observations to a series of model predictions, we must adopt a
parameterization of the {\it local} space density of galaxies for each type, a
cosmological model, and quantify any evolutionary processes (Driver
{\it et al.} 1994). The local space density of galaxies is typically
represented by a
Schechter (1976) luminosity function (LF), which is derived from a local
redshift survey down to some specified magnitude limit.\footnote{The problems
associated with magnitude-limited surveys of this kind are discussed in further
detail in Marzke, Huchra \& Geller (1994) and Driver \& Phillipps (1995).} Two
of the most recent local redshift surveys are those by Loveday {\it et al.}
(1992, LPEM) and Marzke {\it et al.} (1994b, MGHC). They find significantly
different Schechter parameters for the LF of the local populations. Table 3
shows the Schechter parameters derived from these surveys corresponding to the
range of galaxy types adopted in this paper ({\it i.e.} E/S0, Sabc, Sd/Irr).

Parameters for the LPEM-LF, listed in Table 3, were derived as following:
$M_{*}$ and $\alpha$ for E/S0's and Sabc's come directly from the tabulated
LF's for early and late type galaxies. The $\phi_{*}$ values were measured from
Figure 3 of LPEM. As LPEM do not segregate the early-type and late-type
Spirals,
we arbitrarily divide the population by absolute magnitude, assuming galaxies
brighter than $M_{B} = -18$ are Sabc, and those fainter are Sd/Irr
(c.f. Binggeli,
Sandage and Tammann 1988). Hence the Sabc LF is truncated (via an exponential
cut-off, {\it i.e.} $exp[-10^{0.4(M-M^{Cut}}]$) at $M^{Cut} = -18.0$ and
$M_{*} = -18.0$ is adopted for Sd/Irr's. The slope and normalization for
the Sd/Irr's was then chosen such that the (Sabc+Sd/Irr) LF is consistent with
the (Sp/Irr) LF shown in Figure 3 of LPEM. Parameters for the MGHC-LF were
taken as the average between the appropriate classes listed in their Table 1
(as suggested by R. Marzke 1995, private communication). However, note that the
$\phi_{*}$
for the total LF ({\it i.e.} summed over all types) listed in Marzke {\it et
al.} (1995a) is twice that quoted in MGHC and, if correct, would alleviate the
requirement to renormalize the models at $b_{J} = 18$ mag.

Table 3 shows that the level of discrepancy between these two local surveys is
substantial. Hence, for the sake of completeness we shall use both sets of
parameters. The cosmology we adopt is a standard Einstein-de-Sitter model with
$\Lambda=0$, $\Omega=1$, $q_{o}=0.5$, and we also adopt $H_{o}=50$ km s$^{-1}$
Mpc$^{-1}$. K-corrections for galaxy types E/S0, Sa, Sb, Sc, and Sd/Irr were
derived in \S 4. As our main aim is to find any evolution that may have
occurred for each morphological type, we shall assume no-evolution in the model
predictions. To convert the $B$-band Schechter parameters to the $I$-band, we
adopt $(B-I)_{E/S0} = 2.3$ mag, $(B-I)_{Sabc} = 1.9$ mag, and $(B-I)_{Sd/Irr} =
1.4$ mag (from the models in Windhorst {\it et al.} 1994b).

\subsection{The Problem of LF-Normalization in Models}

An additional problem with faint galaxy models is the question of the magnitude
at which to normalize the LF predictions to the observations. The optimal
normalization is done at a flux level where the mean galaxy distance is
sufficiently large that a homogeneous volume is sampled, but not so large that
significant evolution has already taken place. Simply adopting the
normalizations derived from the {\it local} redshift surveys results in a
severe
underestimation of the observed galaxy counts already at relatively low
redshift, where little evolution is expected. This reflects the faster than
Euclidean rise of the observed number counts at bright magnitudes
(Shanks 1989).
Either strong evolution must be occurring locally (Maddox {\it et al.} 1990),
or the local redshift surveys are incomplete for galaxies of dwarf-like
luminosities (Ferguson \& McGaugh 1995), or our location in the Universe is
unusually sparse (e.g. due to large scale structure, etc),\footnote{Note that
if
local space is underdense in galaxies, low-luminosity systems will be severly
under-represented in bright magnitude-limited surveys, because of the smaller
volume over which they are seen (Driver \& Phillipps 1995).} or a combination
thereof. Which of these factors is responsible is not known (see Shanks 1989
for a review).

Figure 8 illustrates the errors associated with faint galaxy models due to this
normalization problem. Figure 8 compares the {\it un}normalized no-evolution
predictions derived from the two sets of Schechter function parameters listed
in
Table 3. Both models under-predict the counts for $m_{b_{J}}\simeq 16$ mag, and
the predictions extrapolated to fainter magnitudes differ even more
significantly from our HST data. Traditionally, most faint galaxy models are
normalized to the observations in the range $18<m_{b_{J}}<22$ mag (Shanks
1989),
which alleviates a good fraction of the discrepancy between the models and
observations at fainter magnitudes. The justification is that at $m_{b_{J}}
\simeq 20.0$ mag even an $0.1 L_{*}$ galaxy will be at a sufficiently large
distance (z$>$0.1; Koo \& Kron 1992) that the sampled volume is likely
homogeneous and isotropic, but not so great that evolutionary processes have
likely already occurred. It is interesting to note that {\it if} the counts are
normalized at $m_{b_{J}}\simeq 18$ mag, the no-evolution models {\it remain} in
good agreement with the observations down to $m_{b_{J}}\simeq 22$ mag, where
the median redshifts is $z\sim$0.25 (KK92). This suggests that we may indeed
live in a sparse {\it local} region of space (for $m_{b_{J}}$\cle 16 mag), and
is difficult to reconcile with a scenario of strong evolution at {\it low}
redshifts (such evolution would then have to have switched off between 18\cle
$m_{B_{J}}$\cle 22 mag, which is rather unlikely). The debate will continue as
to the nature of this discrepancy, so here we will adopt the standard
normalization at $m_{b_{J}}\simeq 18.0$ mag (which is valid until
$m_{B_{J}}\simeq 22$ mag), as illustrated by the filled circle in Figure 8, and
increase the $\phi_{*}$ values listed in Table 3 uniformly for {\it all} galaxy
types by 0.3 dex. We do not mean to imply here that the local surveys are
$\sim$50\% incomplete (although the Zwicky magnitudes may cause some problems
at the faint end of the CfA survey), and emphasize that there is likely a
combination of causes for the normalization problem. A recent LF based on a
nearby galaxy sample selected in the K-band suggested a significantly larger
$\phi_{*}$ value (Glazebrook {\it et al.} 1995) than the optical values
in Table 3.

We mention in this context additional constraints from the well known source
counts and the cosmological evolution of the radio source population (Windhorst
{\it et al.} 1984, 1990, 1993; Condon 1989) on the LF normalization problem.
The
initial steep rise of the radio source counts at Jy levels (with a magnitude
slope of $\sim 0.72$ !) also suggest the presence of a ``local hole'' in the
space density of at least {\it radio} galaxies, and their {\it strong}
cosmological evolution is known {\it not} to start until $z\simeq 0.3$
(Windhorst {\it et al.} 1984, 1990; Condon 1989). Hence, {\it if} field
galaxies had the same space distribution as radio galaxies, LF normalization is
suggested
at $0.1<z<0.2$, consistent with our normalization at $z\simeq 0.15$. This
argument is likely not valid for radio sources with 1.4 GHz fluxes \cge 10 mJy,
which undergo strong cosmological evolution and are {\it not} associated with
field galaxies (Windhorst {\it et al.} 1990), but quite possibly valid for
$\mu$Jy radio sources who also undergo some cosmological evolution (Condon
1989), albeit not as strong and not until z\cge 0.3, and merge into the general
population of field galaxies at the sub-$\mu$Jy level (Windhorst {\it et al.}
1993).

\subsection{Model Predictions for the Morphological HST Counts}

Figure 7 shows the model predictions from the LF's of LPEM and MGHC for each
morphological type with the normalizations adopted in \S 5.3. The LF's of LPEM
and MGHC each suggest different amounts of evolution, when compared to our HST
data in Figure 7. The prediction based on the LPEM-LF (solid lines) appears to
suggest a large amount of evolution in the E/S0's and an inordinate amount of
evolution in the late-type spirals. Alternatively, the MGHC model (dashed
lines)
implies that no-evolution is required in the E/S0 and early-type spirals. LPEM
and MGHC both note that the LPEM-LF is incomplete for early-type galaxies at
the
higher redshifts in the LPEM sample, which could explain the large discrepancy
between the two predictions for E/S0's. {\it If} true, then our HST data for
E/S0's and Sabc's appear to be consistent with the {\it no}-evolution
prediction
of the MGHC-LF (bearing in mind that the increased normalization of the models
at $m_{b_{J}}\simeq 18$ mag $(m_{I}\simeq 16)$ is still unexplained, although
possible reasons are given above). The dashed lines on Figure 7 represent the
locally {\it normalized} predictions.

For late-type spirals/Irregulars, the predictions are drastically different
between the two models. This highlights the large uncertainty with which the
{\it faint} end of the local field LF is known (Driver \& Phillipps 1995), and
how important this part of the LF is in faint galaxy models (Driver
{\it et al.}
1994; Driver 1994). LPEM find a flat faint-end slope $\alpha\simeq -1.0$, and
MGHC find a much steeper slope $\alpha\simeq -1.5$. They actually find a slope
of $\alpha\simeq -1.8$ for the Sm-Im types alone, so in Table 3 we have
averaged the slopes found by MGHC for Sc-Sd and Sm-Im types to generate a
realistic Sd-Im
slope (R. Marzke 1995, private communication). Such a steepening of the faint
end slope is mostly dominated by the late-type population (Binggeli, Sandage \&
Tammann 1988 and MGHC), and has been favored in many recent faint galaxy models
(e.g. Koo \& Kron 1992; Koo, Gronwall \& Bruzual 1993; Driver {\it et al.}
1994; Ferguson \& McGaugh 1995; Phillipps \& Driver 1995). However, these
models typically remain inconsistent with the faint redshift surveys (e.g.
Colless {\it et al.} 1990, 1991, 1993; BES; Cowie, Songalia \& Hu 1991;
Glazebrook {\it et al.} 1995b, and references therein).

The prediction based on an LPEM-LF requires an increase in the galaxy number
density at $m_{I} = 21.75$ of $\sim 1$ dex, while the prediction based on the
MGHC-LF requires $\sim$0.5 dex. To obtain a crude estimate of the amount of
evolution required, we can equate these number density increases to volume
increases, and calculate the magnitude limit of these additional volumes
(assuming that galaxy numbers are preserved). The difference between this
magnitude limit and that of our sample yields an estimate of the amount of {\it
luminosity evolution} required in the {\it entire} population to match the
observed counts ({\it i.e.} $N(m)\propto V(m)\propto L^{\frac{3}{2}}
\Rightarrow \Delta m\propto\frac{1}{0.6}\log[\frac{V_{2}}{V_{1}}]$). This
yields luminosity increases of $\Delta m_{LPEM}\sim 1.7$ and
$\Delta m_{MGHC}\sim 0.8$ mag.

The $M_{*}$ value adopted for the LPEM-Late-type/Irregulars, would
imply that objects with $m_{I} \sim 21.75$ are at z $\cle 0.4$ (or $\cle 6$
Gyrs
in lookback-time). Over this timescale, a typical isolated starburst event will
fade by $\simeq 1.8$ mags (see e.g. Fig. 3 of Wyse 1985 which is based on a
formation starburst within a rapidly condensing gas cloud). However, if there
is a significant underlying population, the {\it total} fading of the {\it
whole} galaxy's luminosity will change by considerably less, dependent on the
luminosity ratio of the new to old populations $\sim$6 Gyrs after the new
burst. The implication is that for a fading of $\Delta m \sim 1.8$, a global
starburst is required with strength comparable to that of the galaxy's initial
formation (or the sum of all previous starbursts). That such events have
occurred in the {\it entire} late-type/Irregular
population over the past 6 Gyrs seems highly improbable, although we note
that the star-formation mechanisms of late-types are poorly known (see Hodge
1989, in which evidence for a wide range of star-forming timescales in local
group members is discussed).

Alternatively, the MGHC-based model,
requires a luminosity evolution of $\Delta m\sim 1.8$, but only
for $\sim$15\% of the population (from consideration of luminosity
conservation). This would then imply that based on a MGHC-based model we
only require a major global starburst event in 15\% of the late-type
poulation over the past 6 Gyrs. Of course other evolutionary scenarioes exist,
but this crude calculation is indicative of the comparative amounts of
evolution required to match the two local LF models to our HST observations.

Figure 9 shows the predicted $B$-band redshift distributions at $m_{I}\simeq
21.75$ mag ($\approx m_{b_{J}}\simeq 23.5$ mag), compared to the faint redshift
surveys of Colless {\it et al.} (1993) and Glazebrook, K., {\it et al.} 1995b.
The predicted redshift distributions have been scaled up to match the observed
distribution. (Note that this is somewhat misleading, as neither LF model
matches the $B$-band number counts at this magnitude, but it is easier to
compare the {\it rescaled} predictions to the {\it shape} of the observed
redshift distribution). The form of the LPEM-LF matches the overall
distribution
closely, while the MGHC-LF both over-predicts the number of low redshift
galaxies and under-predicts the number of high redshift galaxies. Given that
luminosity evolution will shift the peak of the redshift distribution towards
higher redshifts, and that some evolution is required to match the steep
counts,
then the distribution towards lower redshifts of the no-evolution MGHC-LF is
not
unexpected. Of greater concern is the lack of $z>0.5$ objects predicted by MGHC
(due to the low $M_{*}$-values in their E/S0 and Sabc Schechter functions, see
Table 3). The only way to simultaneously reconcile the redshift distribution
and
the morphological counts is through an evolutionary scenario in which the
intrinsically {\it bright} late-types are undergoing evolution at $z\geq 0.5$.
Such models are explored in more detail in Phillipps \& Driver (1995).

\section{The Morphology of Irregular Galaxies in the HST Sample}

Figure 10 shows $V$- and $I$-band grey scale images and contour plots for each
of the 16 Irregulars and the 7 Peculiar types, as described in the caption.
Qualitatively, the majority of this sample is characterized by irregular shaped
outer light-profiles and in many cases a complex nucleus consisting of a number
of higher SB regions. Objects 6, 18, 33, 38, 39, 44, 54, 80, 86, 95, 98, 100,
113, 122, and 126 represent the (15) genuine Irregulars, and objects 29, 65,
73, 74, 83, 104, and 128 represent the (7) "Peculiar" galaxies. The Irregulars
appear to fall into two categories: (1) those which appear inert, and (2) those
which contain multiple cores. Objects 6, 38, 44, 54, 113, and 126 are good
examples of multiple-core objects and comprise about $\sim 40\%$ of the
Irregular sample. Such complex cores may imply active star-formation, e.g. via
spontaneous or merger induced starbursts. Objects 86 and 100 appear to have
strong core structure but in reality this is a manifestation of their much
brighter apparent magnitude. Their apparent central SB is significantly lower
than that of other galaxies of comparable flux (c.f. Table 2). Both galaxies
also show some evidence of merging, but the central regions appear relatively
undisturbed. Objects 54, 80, 122, and 126 also show evidence for a close
companion which {\it may} be responsible for their multiple core structure.
{\it If} a multiple core {\it is} indicative of currently ongoing starformation
({\it
i.e.} evolution), then both spontaneous ({\it i.e.} no obvious merger) and
merger-induced starbursts appear to occur. Nevertheless, half the Irregular
population (objects 18, 33, 39, 86, 95, 98, 100, 129) appears relatively inert
with low SB-cores and with extended irregular shaped light-profiles. This
suggests that {\it both} merger-induced evolution and/or active star-bursts,
{\it and} a higher than anticipated density of local dwarfs contribute to the
faint blue galaxy excess at $m_{I}\simeq 22$ mag. The density of Irregular
galaxies {\it alone} ({\it i.e.} $\sim 50$\% of the Sd/Irr population) appears
to be inconsistent with the LPEM-LF.

If all the ``apparently-evolving'' galaxies ({\it i.e.} $\sim$ 50\%) were
removed from Fig. 7d, this would reduce their observed surface density by {\it
only} 0.3 dex, and our HST data would {\it still} remain inconsistent with the
LPEM-LF model by 0.7 dex ! That is, not only is the LPEM-LF model inconsistent
with our full HST Sd/Irr sample, but it is also inconsistent with the inert
looking HST galaxies alone. Such evidence argues convincingly {\it for} a
MGHC-LF, where the faint end slope of the field LF is steeper ($\alpha$\cge
1.5)
than the often {\it assumed} value ({\it i.e.}, $\alpha\sim$1.0). Hence, it
appears that a combination of evolution and a steep faint end LF is responsible
for the FBG's. The mode of evolution is still unclear, but both mergers and
isolated systems with multiple cores are evident in roughly equal numbers in
our HST images.

\section{Summary and Conclusions}

We presented a complete sample of 144 HST/WFPC2 field galaxies from the Medium
Deep Survey in the magnitude range $20.0\leq m_{I}<22.0$ mag, with the goal of
implementing a full spectroscopic follow-up for all galaxies listed. We have
begun such a long-term program at the MMT (Driver {\it et al.} 1995b). After
classification of these galaxies by eye --- using both the WFPC2
$V$+$I$-morphology and light-profiles --- we compare the global properties of
Irregulars and late-type spirals to those of ellipticals and early-type
spirals.
We find little color difference between early and late-type spirals with mean
$(V-I)$ color's of $1.1\pm 0.2$ mag. The ellipticals have marginally redder
colors with a mean $(V-I) = 1.4\pm 0.2$ mag. An unexpected number of blue E/S0
systems are identified with $(V-I)<1.0$ mag. These could possibly be compact
narrow emission-line galaxies (c.f. Koo {\it et al.} 1995). No Irregulars are
seen with colors redder than $(V-I) = 1.5$ mag. Following the classification
method of Forbes {\it et al.} (1994), we find that the galaxy types can be
reasonably well separated by plotting their Concentration Index [C.I. = core SB
minus total magnitude] versus total magnitude.

We present the differential $I$-band galaxy number counts as a function of
morphological type, excluding those galaxies defined as "Peculiar", and
conclude
that the slopes for the ellipticals and early-type spirals fall at best
marginally above the expected prediction from a {\it no}-evolution models based
on conventional Schechter functions and a standard cosmology. This therefore
implies that strong evolution in the luminous galaxy population is relatively
uncommon down to $m_{I} = 22.0$ mag ({\it i.e.} $z\sim$ 0.5), as also inferred
from the faint galaxy redshift surveys and studies of Lyman-$\alpha$ absorbers
at $z$\cle 1 (Steidel {\it et al.} 1995).

The late-types/Irregular galaxies, however, follow a near-Euclidean slope of
$(\frac{dlog(N)}{dm}) = 0.64\pm 0.1$, indicating either strong evolution in
this
population, local inhomogeneity, and/or a higher than expected {\it local}
space
density of dwarf galaxies, or a combination thereof. From detailed no-evolution
predictions based on the local LF's of LPEM and MGHC, we conclude that a flat
LPEM-LF ($\alpha\simeq -1.0$) is {\it inconsistent} with our HST data, and that
a steep MGHC-LF ($\alpha$\cge 1.5) coupled with a substantial amount of
evolution ($\Delta m\sim 1.8$ mag in only $\sim$15\% of the population) is more
consistent with the data. Examination of the Irregulars alone reveals that
$\sim
40$\% show evidence of interactions or multiple core structure, which suggests
relatively strong and recent evolution in a large fraction of the late-type
population. This work will continue with the study of a deeper field from a
24-orbit HST exposure (Driver {\it et al.} 1995a), and in a systematic
spectroscopic follow-up of the entire current sample (Driver {\it et al.}
1995b).

\bigskip

We would like to acknowledge the work of other members of the Johns Hopkins
University Medium Deep Survey team (Kavan Ratnatunga, Lyman Neuschaefer, Eric
Wyckoff and Stefano Casertano) in the initial data reduction stage. We would
also like to thank Roger Rouse for his help in the galaxy classification
process, Dave Burstein for useful and informative discussions, Karl Glazebrook
for making his independent MDS galaxy catalog available, Myungshin Im for help
with the K-corrections, and the anonymous referee for constructive comments on
the original manuscript. SPD, RAW and REG acknowledge financial support from
HST MDS grant GO.2684.03.93A(ASU) and .01.93A(JHU).

\pagebreak

\section*{References}

\begin{description}

\baselineskip=18pt

\item Babul, A., \& Rees, M. J. 1992, MNRAS, { 255}, 346

\item Bahcall, J. N., Flynn, C., Gould, A., \& Kirhakos, S. 1994, ApJL, {
435}, L51

\item Binggeli, B., Sandage, A., \& Tammann, G. A. 1988, ARA\&A, { 26}, 509

\item Broadhurst, T. J., Ellis, R. S., \& Shanks, T. 1988, MNRAS, { 235}, 827
(BES)

\item Broadhurst, T. J., Ellis, R. S., \& Glazebrook K. 1992, Nature, { 355},
827

\item Burkey, J. M., Keel, W. C., Windhorst, R. A., \& Franklin, B. E. 1994,
ApJL, { 429}, L13

\item Burstein, D., \& Heiles, C. 1982, ApJ, { 87}, 1165

\item Casertano, S., Ratnatunga, K. U., Griffiths, R. E. , Im, M., Neuschaefer,
L. W., Ostrander, E. J., \& Windhorst, R. A. 1995, ApJ, submitted (CRGINOW).

\item Cawson, M. G. M. 1983, PhD thesis, University of Cambridge

\item Charlot, S., \& Bruzual, G. 1991, ApJ, { 367}, 126

\item Colless, M., Ellis, R. S., Taylor, K., \& Hook, R. N. 1990, MNRAS, {
244}, 408

\item Colless, M., Ellis, R. S., Taylor, K., \& Shaw, G. 1991, MNRAS, { 253},
686

\item Colless, M., Ellis, R. S., Broadhurst, T. J., Taylor, K., \& Peterson, B.
A. 1993, MNRAS, { 261}, 19

\item Colless, M., Schade, D., Broadhurst, T. J., \& Ellis, R. S. 1994, MNRAS,
{ 267}, 1108

\item Condon, J. J. 1989, ApJ, { 338}, 13

\item Cowie, L. L., Songalia, A., \& Hu, E. M. 1991, Nature, { 354}, 460

\item Disney, M. J. 1976, Nature, { 263}, 573

\item Driver, S. P. 1994, PhD Thesis, University of Wales

\item Driver, S. P., Phillipps, S., Davies, J. I., Morgan, I., \& Disney M. J.
1994, MNRAS, { 266}, 155

\item Driver, S. P., \& Phillipps, S. 1995, ApJ, submitted

\item Driver, S. P., Windhorst, R. A., Ostrander, E. J., Keel, W. C.,
Griffiths, R. E., \& Ratnatunga, K. U. 1995a, ApJL, submitted

\item Driver, S. P., Windhorst, R. A., Pascarelle S. M., Griffiths R. E. 1995b,
in preparation

\item Ferguson, H. C., \& McGaugh, S. S. 1995, ApJ, { 440}, 470

\item Forbes, D. A., Elson, R. A. W., Illingworth, G. D., \& Koo, D. C. 1994,
ApJL, { 437}, L17

\item Giraud, E. 1992, A \& A, { 257}, 501

\item Glazebrook, K., Ellis, R. E. Santiago, B. \& Griffiths, R. E. 1995a,
MNRAS, submitted (GL95a)

\item Glazebrook, K., {\it et al.} 1995b, MNRAS, { 273}, 157

\item Glazebrook, K., Peacock, J. A. Miller, L. \& Collins, C.A. 1995c, MNRAS,
submitted

\item Griffiths, R. E. {\it et al.} 1994a, ApJL, { 435}, L49 (GR94a)

\item Griffiths, R. E. {\it et al.} 1994b, ApJ, { 437}, 67 (GR94b)

\item Guiderdoni, B., \& Rocca-Volmerange, B. 1987, A\&A, { 186}, 1

\item Guiderdoni, B., \& Rocca-Volmerange, B. 1988, A\&AS, { 74}, 185

\item Guiderdoni, B., \& Rocca-Volmerange, B. 1991, A\&A, { 252}, 435

\item Hodge, P. 1989, ARA\&A, { 27}, 139

\item Holtzman, J. A., {\it et al.} 1995, PASP, { 107}, 156

\item Im, M., Casertano, S., Griffiths, R. E., Ratnatunga, K. U. \& Tyson,
J. A. 1995a, ApJ, { 441}, 494

\item Im, M. {\it et al.} 1995b, ApJ, in press

\item Jones, J. B., Driver, S. P., Phillipps, S., Davies, J. I., Morgan, I.,
\& Disney, M. J. 1995, in preparation

\item Keel, W. C., \& Windhorst, R. A. 1993, AJ, { 106}, 455

\item Kent, S. M. 1985, ApJS, { 59}, 115

\item Koo, D. C., \& Kron, R. G. 1992, ARA\&A, { 30}, 613 (KK92)

\item Koo, D. C. Guzman, R., Faber, S. M., Illingworth, G. D., \& Bershady, M.
A. 1995, ApJL, { 440}, L49

\item Koo, D. C., Gronwall, C., \& Bruzual A., G. 1993, ApJL, { 415}, L21

\item Kron, R. G. 1980, ApJS, { 43}, 305

\item Lacey, C. G. 1991, Nature, { 354}, 458

\item Lacey, C. G., Guiderdoni B., Rocca-Volmerange B., \& Silk J. 1993, ApJ,
{ 402}, 15

\item Lilly, S. J., Cowie, L. L., \& Gardner, J. P. 1991, ApJ, { 369}, 79

\item Loveday, J., Peterson, B. A., Efstathiou, G. \& Maddox, S. J. 1992, ApJ,
{ 390}, 338 (LPEM)

\item Maddox, S. J., Sutherland, W. J., Efstathiou, G., \& Loveday, J. 1990b,
MNRAS, { 243}, 692

\item Marzke, R. O., Huchra, J. P., \& Geller, M. J. 1994a, ApJ, { 428}, 43

\item Marzke, R. O., Geller, M. J., Huchra, J. P., \& Corwin, H. G. 1994b, AJ,
{ 108}, 437 (MGHC)

\item McGaugh, S. S. 1994, Nature, { 367}, 538

\item Mutz, S. B., {\it et al.} 1994, ApJL, { 434}, L55

\item Neuschaefer, L. W., \& Windhorst, R. A. 1995, ApJ Sup. { 96}, 371

\item Neuschaefer, L. W., Griffiths, R. E., Ratnatunga, K. U., \& Valdes, F.
1995, PASP, in press

\item Osterbrock, D. E. 1989, {\it Astrophysics of Gaseous Nebulae and Active
Galactic Nuclei}, University Science Books

\item Phillipps, S., \& Driver, S. P. 1995, MNRAS, in press

\item Ratnatunga, K. U., Griffiths, R. E., \& Casertano, S. 1994, in {\it The
Restoration of HST images and Spectra}, eds Hanish, R. J., \& White, R., Proc.
STScI Workshop. p. 333.

\item Ratnatunga, K. U., Griffiths, R. E., Casertano, S., Neuschaefer, L. W.,
\& Wyckoff, E. W. 1995a, ApJ, submitted

\item Rocca-Volmerange, B., \& Guiderdoni, B. 1990, MNRAS, { 247}, 166

\item Sandage A. 1961, {\it The Hubble Atlas of Galaxies}, Carnegie Institution
of Washington Publ. No 618 (Washington DC)

\item Schechter, P. 1976, ApJ, { 203}, 297

\item Shanks, T., Stevenson, P. R. F., Fong, R. \& MacGillivray, H. T. 1984,
MNRAS, { 206}, 767

\item Shanks, T. 1989, in {\it The Extra-galactic Background Light}, eds
Bowyer, S. C., \& Leinert, C. publ Kluwer Academic Publishers

\item Steidel, C. C., Dickinson, M., Persson, S. E. 1994, ApJL, { 437}, L75

\item Tyson, J. A. 1988, AJ, { 96}, 1

\item Windhorst, R. A. 1984, PhD Thesis, Univ. Leiden

\item Windhorst, R. A., Mathis, D. F., \& Neuschaefer, L. W. 1990, in ASP Conf.
Ser., Vol. 10, Evolution of the Universe of Galaxies (Edwin Hubble Centennial
Symposium), ed. R. G. Kron (Provo, UT: BookCrafters, Inc.), 389--403

\item Windhorst, R. A., Fomalont, E. B., Partridge, R. B., \& Lowenthal, J. D.
1993, ApJ, { 405}, 498

\item Windhorst, R. A., et al. 1994a, AJ, { 107}, 930

\item Windhorst, R. A., Gordon, J. M., Pascarelle, S. M., Schmidtke, P. C.,
Keel, W. C., Burkey, J. M., \& Dunlop, J. S. 1994b, ApJ, { 435}, 577

\item Windhorst, R. A., Franklin, B. E., \& Neuschaefer, L. W. 1994c, PASP,
{106}, 798

\item Wyse, R. F. 1985, ApJ, { 299}, 593

\end{description}

\pagebreak

\section*{Tables}

\begin{center}

Table 1a --- Field positions and exposures times for selected MDS fields.

\medskip

\begin{tabular}{lccrrrrrrrr}
\hline
\hline
\n Field &RA&DEC&l$^{II}$&b$^{II}$&\multicolumn{2}{c}{Exposure time}&
\multicolumn{2}{c}{No of Orbits}&$A_{V_{606}}$&$A_{I_{814}}$\\
\n Name  &\multicolumn{2}{c}{(J2000)}&\multicolumn{2}{c}{(deg)}&V&I&V&I&
\multicolumn{2}{c}{(mag)}\\
\hline
\smallskip
\n ueh0 & 00:53:23.2 & +12:33:58 &123.68 &--50.30& 8700 & 6300 & 5 & 3 & 0.13
& 0.10\\
\n uim0 & 03:55:31.6 & +09:43:34 &179.83 &--32.15& 8800 & 5200 &12 & 6 & 0.31
& 0.23\\
\n uop0 & 07:50:47.1 & +14:40:44 &206.07 & 19.63 & 7200 & 4200 & 5 & 2 & 0.04
& 0.03\\
\n usa2 & 17:12:23.2 & +33.35:49 & 56.72 & 34.25 & 5400 & 6300 & 3 & 3 & 0.08
& 0.06\\
\n ux40 & 15:19:41.2 & +23:52:06 & 35.78 & 56.51 & 3300 & 7500 & 2 & 4 & 0.10
& 0.07\\
\n uy40 & 14:34:48.7 & +25:08:02 & 33.87 & 66.75 & 5200 & 6000 & 6 & 6 & 0.06
& 0.04\\
\hline
\end{tabular}

\end{center}

\vspace{1.50cm}

\begin{center}

Table 1b --- Statistics and detections for the selected MDS fields.

\medskip

\begin{tabular}{crrrrccc}
\hline
\hline
\n Field &$K_{V}$ &$K_{I}$&$2\sigma \mu_{V}$&$2\sigma \mu_{I}$&
\multicolumn{3}{c}{No of Detections} \\
\n       &        &       &                 &                 & WFC2 & WFC3
& WFC4     \\
\hline
\smallskip
\n  ueh0i & 33.23 & 32.45 & 23.8            & 23.4            &  7   & 11
&  7       \\
\n  uim0i & 32.53 & 31.48 & 24.2            & 23.4            &  8   &  6
&  5       \\
\n  uop0i & 33.09 & 32.45 & 24.3            & 23.3            &  5   & 10
& 12       \\
\n  usa2i & 33.33 & 32.45 & 25.0            & 24.3            &  6   &  7
& 14       \\
\n  ux40i & 33.23 & 32.32 & 23.5            & 24.1            &  9   &  6
&  9       \\
\n  uy40i & 32.53 & 31.64 & 24.7            & 23.9            &  7   & 12
&  3       \\
\hline
\n  Total &       &       &                 &                 &
\multicolumn{3}{c}{144} \\
\end{tabular}

\end{center}

\smallskip

\n {\it Note:} $K_{V}$ and $K_{I}$ are the calibration constants for the $V$
and $I$ fields, respectively, with the exposure time and a scaling factor
incorporated ({\it i.e.} observed mag = K -- 2.5 log ADU$_{\mbox{tot}}$).

Table 2 --- See http://www.phys.unsw.edu.au/~spd/bib.html

Table 3 --- Schechter function parameters for the LF's of different galaxy
classes.

\medskip

\begin{center}

\begin{tabular}{llccc}
\hline
\hline
\n Survey & Type   & $M_{*}$ & $\alpha$ & $\phi_{*}$            \\ 
\hline
\smallskip
\n LPEM   & E/S0   & --21.2  & +0.2     & $4.00 \times 10^{-4}$ \\
\n        & Sabc   & --20.9  &--0.8     & $1.00 \times 10^{-3}$ \\
\n        & Sd/Irr & --18.5  &--1.1     & $7.00 \times 10^{-4}$ \\
\n MGHC   & E/S0   & --20.5  &--0.9     & $1.14 \times 10^{-3}$ \\
\n        & Sabc   & --20.3  &--0.8     & $1.74 \times 10^{-3}$ \\
\n        & Sd/Irr & --20.3  &--1.5     & $2.50 \times 10^{-4}$ \\
\hline
\end{tabular}

\end{center}

\pagebreak

\section*{Figures}

\bigskip

\n {\bf Figure 1a (upper panel)} --- Signal-to-noise ratio versus magnitude for
the {\it isophotal} detection stage. {\bf Fig 1b (lower panel)}, as Fig 1a, but
for the {\it aperture} photometry stage. The horizontal line represents the
minimum detectable signal-to-noise ratio from the detection criterion, and the
vertical lines represents the selected magnitude range. Only those galaxies
included in the final sample are shown here. The dashed line is the
non-cosmological relation between signal-to-noise ratio versus aperture
magnitude expected for a single non-evolving galaxy.

\bigskip

\n {\bf Figure 2} --- A selection of greyscale plots and major-axis
light-profiles used for the eyeball classifications. The greyscale plots are
plotted as log(Intensity) from $\mu_{I} = 21.0$ mag/arcsec$^{2}$ to sky.
Major-axis profiles are shown as SB vs. radius (r) and SB vs.
$r^{\frac{1}{4}}$.
A representative galaxy for each morphological class is shown at comparable
magnitudes ($m_{I} \sim 21.5$ mag). Only the $I$-band data is shown here, but
equivalent $V$-band data was also used in the classification process.

\bigskip

\n {\bf Figure 3} --- A comparison between the morphological classification and
the concentration of a galaxy's light. The concentration index (C.I.) is the
apparent central SB or core aperture magnitude minus the total magnitude. A low
value of the C.I. suggests that the majority of the light is concentrated
towards the center. The ellipticals form a clearly distinct population, but the
late-types are less well distinguished by this method, as expected.

\bigskip

\n {\bf Figure 4} --- The $(V-I)$ color for the sample shows little or no trend
with apparent I-band magnitude or morphological type, although few Irregulars
and late-types are seen with colors redder than $(V-I) =1.5$. Note the
non-negligible fraction of blue galaxies classified as E/S0 which have
r$^{1/4}$
light-profiles. Also shown is the histogram of the $(V-I)$ data for the entire
sample and for the Irregulars alone (shaded area).

\bigskip

\n {\bf Figure 5} --- The expected relation between color and redshift for the
Hubble sequence, using no-evolution, but simple K-corrections derived from the
present day spectra of Guiderdoni \& Rocca-Volmerange (1987, 1988) for the
WFPC2 $V_{606}$ and $I_{814}$ filters (Myungshin Im 1995, private
communication).

\bigskip

\n {\bf Figure 6} --- The number of galaxies observed for each type as a
function of $I$-band magnitude. This shows a rapid increase in the numbers of
Irregular galaxies over the conventional spirals and ellipticals. The initial
mix of 36\% E, 50\% Sabc and 14\% Sd/Irr at $m_{I}=20.25$ mag becomes 28\% E,
35\% Sabc and 31\% Sd/Irr at $m_{I}=21.75$ mag. Typical errors in these
percentages are $\sim 5-8$\%. If this trend continues, then late-types and
Irregulars will make up the bulk of the galaxy population observed at fainter
magnitudes.

\bigskip

\n {\bf Figure 7} --- Differential galaxy number counts for: (a) our complete
HST sample; (b) for E/S0 galaxies; (c) for Sabc galaxies; and (d) for late-type
galaxies (Sd/Irr). The slope of the overall counts is consistent with data from
other groups (see Fig. 8) and implies that our sample is complete at least down
to I=22.0 mag. The ellipticals and spirals (panels b and c) have shallower
slopes. The model predictions are for a LPEM-LF model (solid lines) and a
MGHC-LF model (dashed lines). For panels b and c, the MGHC-LF models agree well
with our observed HST counts using no-evolution and a standard cosmology.
However, the late-type galaxies follow a much steeper slope, indicating that
either they are more local and less affected by cosmology, and/or that they are
undergoing a significant amount of evolution. The model lines are assuming two
alternate luminosity functions and no-evolution.


\bigskip

\n {\bf Figure 8} --- The differential galaxy number counts in the $b_{J}$-band
from various sources listed in the figure. The model lines show predictions
using the {\it non-evolving} LPEM-LF (solid line) and the MGHC-LF (small dashed
line). These model lines are {\it un}normalized ({\it i.e.} as locally
determined). Not only do the two models give dramatically different predictions
at faint magnitudes, but also does neither model match the counts fainter than
$m_{b_{J}} \simeq 17$ mag, suggesting local evolution, local selection effects,
and/or local inhomogeneity (see text). Most galaxy models are normalized to the
counts at $18.0<m_{b_{J}}<22.0$ mag (c.f. Shanks 1989, see \S 5.2). Here we
normalize at $m_{b_{J}} \simeq 18$ mag, as indicated by the single solid
circle,
which implies a 0.3 dex increase of the local LF. The {\it normalized}
Marzke-LF model is shown as a large dashed line.

\bigskip

\n {\bf Figure 9} --- The predicted redshift distribution for
$m_{b_{J}} = 23.5$
mag compared to the data of Colless {\it et al.} (1993) and Glazebrook {\it et
al.} (1995). The model distributions are scaled to reflect the discrepancy
between the observed and predicted number counts at $m_{I} \sim 23.5$ mag. The
shaded areas represent the contribution from the late-type/Irregulars
population.

\bigskip

\n {\bf Figure 10} --- For each of the 16 Irregular and 7 Peculiar galaxies a
section of six panels are shown. Each panel represents a 5$\times$5'' box. The
panel columns represent: the $I$-band image; a contour plot of the $I$-band
image smoothed with a 0.2'' FWHM Gaussian; the smoothed $I$-band image; and the
corresponding plots in the $V$-band. For the first and third columns the data
is displayed from $I\mu$=22.0 to $I\mu$=25.0 mag/arcsec$^{2}$. In the second
column, each contour represents an increase in 0.5 mag/arcsec$^{2}$ in SB
starting at $I\mu$=24.5 mag/arcsec$^{2}$. The fourth and sixth columns are
displayed from $V\mu$=23.0 mag/arcsec$^{2}$ to sky, and the contours for the
fifth column are again 0.5 mag/arcsec$^{2}$ intervals in SB, but starting at
$V\mu=25.5$ mag/arcsec$^{-2}$.

\end{document}